\documentstyle[prl,aps,epsf,twocolumn]{revtex}
\tighten
\begin{document}

\title{ Origin of strong scarring of 
wavefunctions in quantum wells in a tilted
magnetic field}   
\author{E.~E.~Narimanov   and  A.~Douglas  ~Stone}  
\address{Applied  Physics,  
Yale  University,  P.O.  Box  208284,  New Haven CT
06520-8284} 
\date{\today} 
\maketitle 

\begin{abstract} 
The anomalously strong scarring of wavefunctions
found in numerical studies of quantum wells in a tilted magnetic field
is shown to be due to special properties of the classical dynamics
of this system.  A certain subset of periodic orbits are identified
which are nearly stable over a very large interval of variation of
the classical dynamics; only this subset are found to exhibit strong
scarring.  Semiclassical arguments shed further light on why these
orbits dominate the experimentally observed tunneling spectra.
\end{abstract}   

\narrowtext
The localization of certain quantum wavefunctions in real-space along
unstable classical periodic orbits illustrates how quantum mechanics
can violate the ergodic behavior expected from classical mechanics.
Such wavefunctions are conventionally termed ``scars'' \cite{Heller}
and their properties have been extensively studied by theorists of
quantum chaos during the past decade \cite{Bogomolny,Berry,Tomsovic}.  
Recently, an
experimental system has been discovered and studied 
\cite{Fromhold,Muller} in which such scarred wavefunctions
control to a large extent an observable physical property, the
tunneling current through a double-barrier GaAs-AlGaAs heterostructure
(``quantum well'') under high bias.  When a magnetic field is applied 
at an angle $\theta$ with respect to the normal to the barriers 
(the electric field direction), the resulting dynamics makes a
transition to chaos \cite{Fromhold,Muller,ss} as $\theta$ is increased
from zero.  Calculations by Fromhold et al. \cite{scars} on
the system found many more scars than in any previously-studied
quantum-chaotic hamiltonian, {\it and} that these scarred wavefunctions
carried most of the the tunnel current when the system was resonant.
In the initial experiments \cite{Fromhold,Muller} the level-broadening
(due to optic phonon emission) was too large to observe the resonances
due to individual levels. However in a later experiment \cite{Nature}
this was done, albeit at such low quantum numbers that the concept of
scarring becomes less meaningful \cite{Monteiro2}.  From extensive
numerical \cite{Nature,scars,Monteiro3,long} work we know: 
1) Quantum states scarred by the same periodic 
orbit appear over a wide
range of variation of the classical dynamics, in contrast to typical
systems (e.g. billiards).  2) The scars arise
from only a few of the many unstable short periodic orbits in the tilted
well.  3) These scars carry most of the resonant tunneling current for
$\theta > 15^{\circ}$.  In this Letter we will present a theory to
explain why only certain orbits scar the wavefunctions and why these 
scars persist as the classical dynamics changes substantially.
The theory also sheds further light on why
these scarred states dominate the tunneling current.

First, we recall why in typical chaotic systems scarred states are
relatively rare.  Generally scarred states are associated with orbits
which are not too unstable; a sufficient condition for strong 
scarring is that
$\lambda T \leq 1$, where $\lambda$ is the largest instability 
(Lyapunov) exponent associated with the orbit.  Typically in chaotic
hamiltonian systems, unstable periodic orbits appear as marginally stable
orbits at bifurcations and become monotonically more unstable as the
classical parameter (e.g. energy) driving the system to chaos is increased.
Therefore such orbits only scar over the small interval of classical
parameter space when they are close to stability.  In a recent detailed
study of the classical dynamics of the tilted well \cite{long}, we have
shown that in this system a subset of the short periodic orbits behave 
completely differently.  They exist only for a finite interval as the 
classical parameters (energy $\varepsilon$, magnetic field, $B$ and electric
field $E$) are varied and are ``pinned'' near marginal stability for
their entire interval of existence.  Therefore these orbits are
responsible for the strong scarring seen, and indeed all the
scars found numerically correspond to this subset of the 
short orbits.  We now present this argument in detail.

We model the system by two infinite potential barriers corresponding to
the $x-y$ planes at $z=-d$ (the emitter) and $z=0$ (the collector), 
with an electric field ${\bf E}=
E \hat{z}$, and a ``tilted'' magnetic field ${\bf B}= B \cos \theta 
\hat{{\bf z}}
+ B \sin \theta \hat{{\bf y}}$ for $-d<z<0$.  The periodic orbit theory for
this model is rather involved \cite{long} and we only sketch the most
salient features here.  First, the classical hamiltonian can be rescaled
\cite{long} so that the dynamics only depends on two dimensionless
parameters: $\beta = 2Bv_0/E$ and $\gamma = \varepsilon/eV$ where $V$
is the voltage across the well and $v_0=(2 \varepsilon /m^*)^{1/2}$ is
the velocity corresponding to the total injection energy.  In the
experiments $\gamma \approx 1.17$ is constant to a good approximation 
\cite{long} and so $\beta$, the scaled magnetic field, 
is the single relevant variable.  Our semiclassical
analysis \cite{long} expresses the tunneling current in terms of  
periodic orbits in the well which connect the emitter and collector
barriers (we term these ``emitter'' orbits, those which don't reach
the emitter ``collector'' orbits).  We focus on
the experimentally relevant situation in which $\gamma$ is only slightly
greater than unity, so that many of the periodic orbits are collector
orbits and may be neglected.  Optic phonon emission
produces a temporal cut-off which allows only short periodic
orbits to produce structure in the tunneling spectra. 

Emitter orbits have the following properties \cite{long}.  An orbit
which collides with the collector $n$ times (period-$n$ in the collector
Poincar\'e map) can collide with the emitter $m$ times, where $m=1,2,
\ldots, n$.  Hence we denote the orbits as $(m,n)$. All emitter orbits
(except for a single $(1,1)$ ``traversing'' orbit) only
exist above a threshold value of $\beta = \beta_{c1}$ and cease to
exist at a higher value of $\beta = \beta_{c2}$ by an inverse bifurcation
(usually a tangent bifurcation (TB)).  While the ``death'' of the 
orbits follow the generic rules of hamiltonian bifurcation theory 
\cite{Meyer}, {\it all} relevant orbits are ``born'' in a new kind 
of bifurcation,
which we refer to as a cusp bifurcation (CB).  CB's have non-generic 
properties since they appear on a closed curve in the surface of section
(SOS), where the 
the Poincar\'e map describing the dynamics 
is nonanalytic. This ``critical'' curve separates initial conditions at the 
collector barrier which will reach the emitter on the next try 
from those which will not. Hence orbits originating just within the curve 
will receive a
``kick'' at the emitter, while those just outside will not.  This
leads to a discontinuity in the stability matrix $M$ of any
periodic orbit coresponding to a fixed point which crosses the boundary.  
In particular, all CB's occur
by the simultaneous appearance of two new orbits, which infinitesimally
above threshold differ by one point of contact with the emitter (this
may correspond to either one or two fewer collisions).
We have shown \cite{long} 
that the orbit which reaches the emitter more times
has diverging stability ($|Tr M | \to \infty$) at $\beta_{c1}$ while the
other orbit in the pair can be either stable ($|Tr M | < 2$) or unstable 
($|Tr M | > 2$) at $\beta_{c1}$. The latter case, in which two
{\it unstable} orbits appear simultaneously is forbidden for generic
hamiltonian systems \cite{Meyer}.  Continuity arguments and numerical results 
\cite{long}
imply that the partner in a cusp bifurcation with fewer emitter collisions 
will have $|Tr M | \sim 2$, i.e. will be born near marginal stability.
Hence such orbits, being born near marginal stability and being required
to return to marginal stability when they disappear in 
an inverse tangent bifurcation, remain
only slightly unstable for their entire interval of existence, {\it unlike}
unstable periodic orbits of typical chaotic systems.  It is this subset
of the orbits which scar strongly the quantum states over a large
variation of the classical parameter $\beta$.

Bifurcation and stability diagrams illustrating this behavior are shown
in Fig. 1 for the case of period-two and period-three orbits. 
Among the period-$2$ orbits the orbit denoted $(1,2)_2$ fits our criteria.  
At $\theta = 28^\circ$ it is born in a CB
with the higher connectivity orbit $(2,2)^+$ at $\beta \simeq 4.0$ 
and dies in an inverse TB with the orbit $(0,2)^-$ at $\beta \simeq 7.0$.
This period-two orbit and another topologically similar orbit 
(not shown) which
appears at slightly higher value of $\beta$ account very well 
\cite{long} for the 
peak-doubling regions observed at $\theta = 28^{\circ}$ in the
experiments of Muller et al. \cite{Muller}.  This orbit
was found to scar
many wavefunctions in the work of Fromhold et al. \cite{Nature}.
It has a complicated evolution above $\theta = 29^{\circ}$, involving
an ``exchange'' bifurcation with the topologically similar orbit just
mentioned, the details of which are given elsewhere \cite{long}.
Note however that in roughly the same $\beta$ interval there are 
two other period-two emitter orbits, each with rapidly varying
stability.  Both born in CBs paired with an orbit with fewer
collisions with the emitter, hence by our above reasoning
are initially enormously unstable. Therefore they do not
generate strong scars in the spectrum.

A similar story holds for one of the eight period-three orbits
which appear around $\beta=3.5$ at $\theta=38^{\circ}$, 
the orbit we denote $(1,3)^-$ (Fig. 1). 
This orbit has been discussed previously 
\cite{Fromprb,Fromcomment,Frombig} in
connection with the observability of trifurcations in the data of
ref. \cite{Muller}.  It is born in a CB as the
partner of a $(3,3)$ orbit, remains
near marginal stability for $3.2 < \beta < 4.4$  and dies in a TB with
the $(1,3)^+$.  In the same interval there are several other unstable 
period-three emitter orbits which do not scar strongly.
Finally, by the same reasoning we have 
found a $(1,5)$ orbit which scars strongly.  Quantum states 
scarred by each of these orbits are shown in Fig. 2.

Note that by our criteria the scarring orbits must always be
$(m,n)$ orbits with $m < n$; e.g. (1,2) can scar strongly
whereas (2,2) should not.  On the other hand, it is easily shown
\cite{long} that as $\theta \to 0$ the {\it only} emitter orbits are of
the type $(n,n)$.  Therefore the interval of existence of the scarring
orbits is small for small $\theta$.  Thus, for example, 
the period-three scarred states are unimportant for $\theta < 20^{\circ}$.

We have tested this argument quantitatively 
by analyzing the quantum states of the
tilted well for scars of the three orbits shown in Fig. 2.
By generating many spectra at different values of $B,E$ we can search
in the experimentally appropriate intervals of $\beta$ with 
$1.1 < \gamma <1.2$ and systematically detect these scars.  In Fig. 3 we plot
a measure of scar strength versus action of the scarring orbit.
As noted before \cite{Agam,scars}, the energies $\epsilon_n$ of scarred states 
satisfy an approximate Bohr-Sommerfeld quantization rule, $S(\epsilon_n)
= (n + \phi)2 \pi \hbar$, so we expect and find a strong periodic modulation 
which turns on at the tangent bifurcation at which the orbit appears 
\cite{ghost}.

Finally, we comment on the fact that these scarred states tend to dominate
the tunneling current at large tilt angles.  All of the orbits studied
here and elsewhere \cite{scars,Nature,Monteiro3} which scar strongly
have only a single collision
with the emitter barrier.  Since both emitter and collector surfaces
of section must be symmetric when $v_y \to -v_y$, such orbits must
have $v_y=0$ at the emitter barrier \cite{pitchfork}.  
Thus these orbits have unusually
low transverse velocity at the emitter barrier compared to other 
orbits with the same periodicity in the collector Poincar\'e map.
Since the emitter wavefunction is primarily a superposition of the 
first few 
Landau levels, the source of tunneling current has low transverse
velocity and couples very well to these scarred states.
Therefore it is due to this specific feature of the emitter state 
that these scarred well states dominate the tunneling current; if the
emitter state had large transverse momentum these states would be
anti-selected.  Since scarred states are localized in
real and phase space, they can lead to quasi-selection rules for
tunneling, but only if they are localized in the correct regions
to couple well to the input state.

We acknowledge helpful conversations with Greg Boebinger,
Mark Fromhold and Tania Monteiro and the
support of NSF grant DMR-9215065.

\onecolumn

\begin{figure}[htbp]
\begin{center}
\leavevmode
\epsfysize=5.in 
\epsfbox{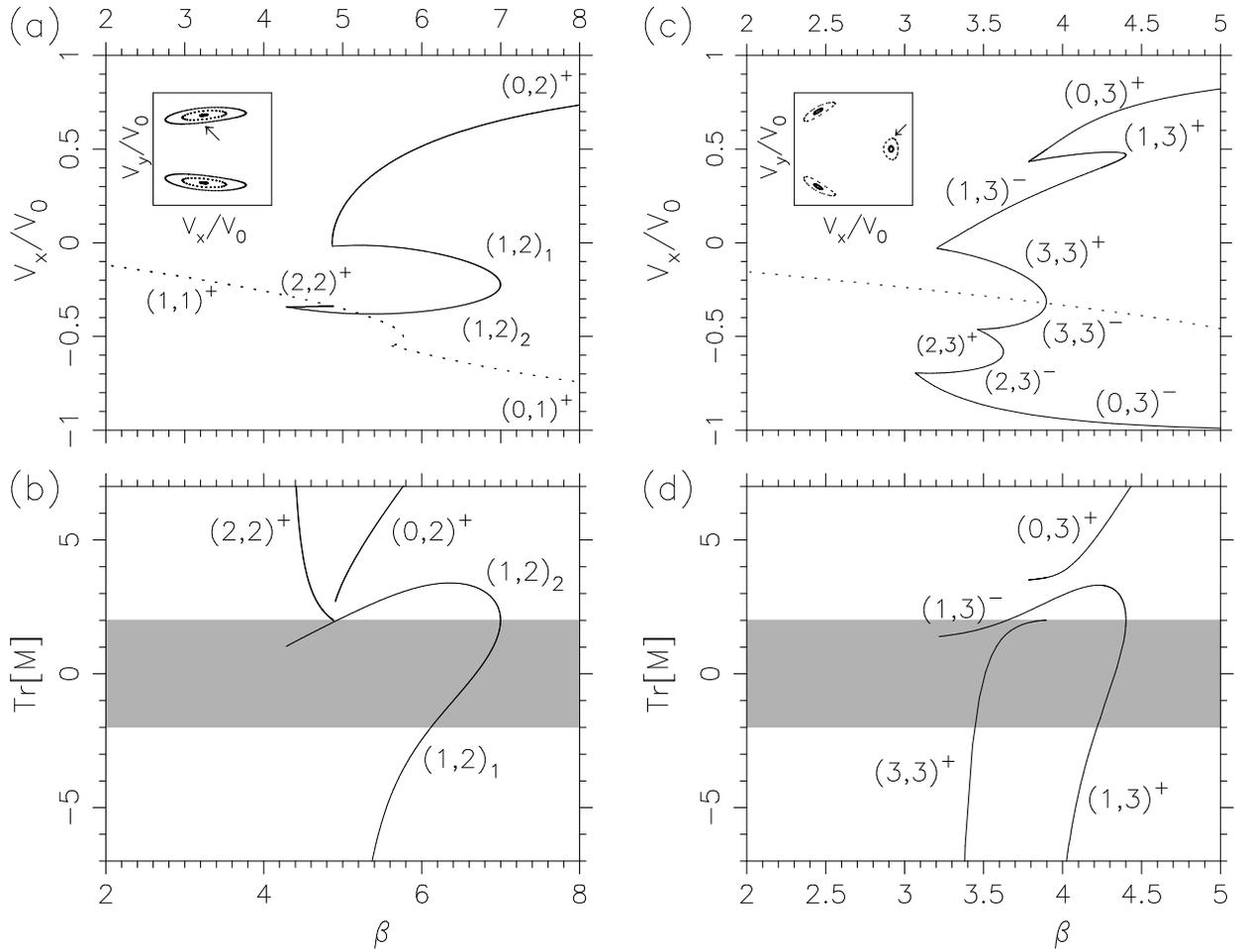}
\end{center}
\caption{ 
Bifurcation diagrams for the relevant period-two (a) and period-three 
(c) orbits.  The horizontal axis is the scaled magnetic field, $\beta$,
the vertical axis is the $v_x$ coordinate of the fixed point(s) in the 
collector SOS indicated schematically in the insets.  
(b) and (d) are plots of the
trace of the monodromy (stability) matrix for the corresponding orbits;
Shaded area ($|{\rm Tr[M]}| < 2$) denotes stability region.  Orbits
denoted $(1,2)_2,(1,3)^-$ (see Fig. 2) remain slightly unstable over 
a large variation of $\beta$, leading to strong scarring.
\label{fig_po2_bd} }
\end{figure}

\begin{figure}[htbp]
\begin{center}
\leavevmode
\epsfxsize=7.in 
\epsfbox{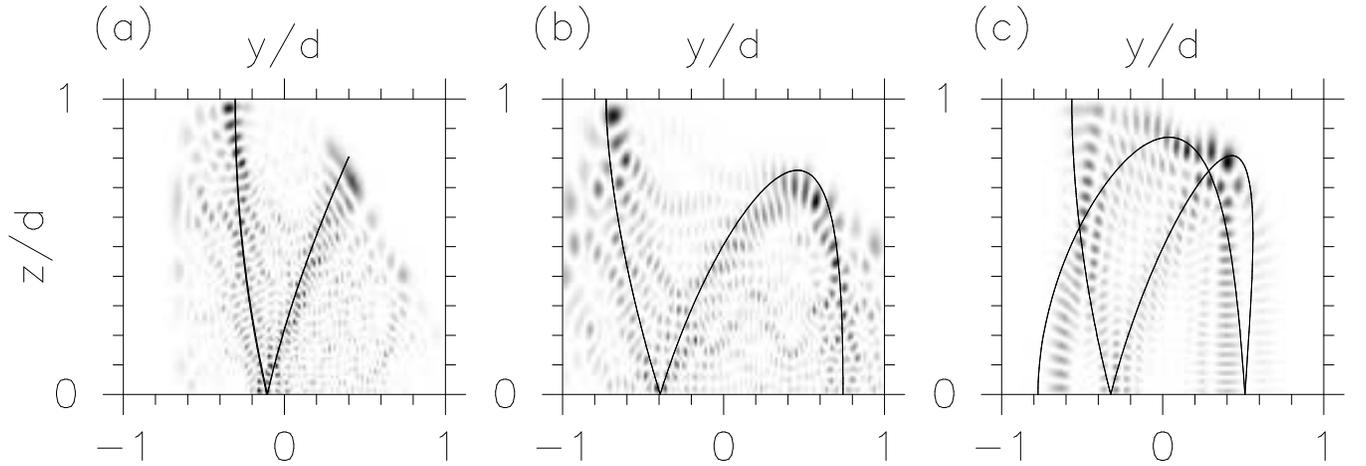}
\end{center}
\caption{
Examples of wavefunctions scarred by the unstable
$(1,2)_2$ orbit (a), $(1,3)^-$ orbit
(b) and a $(1,5)$ orbit (c); y-z projections of orbits are superimposed.
\label{psi} 
}
\end{figure}

\begin{figure}[htbp]
\begin{center}
\leavevmode
\epsfysize=2.5in 
\epsfbox{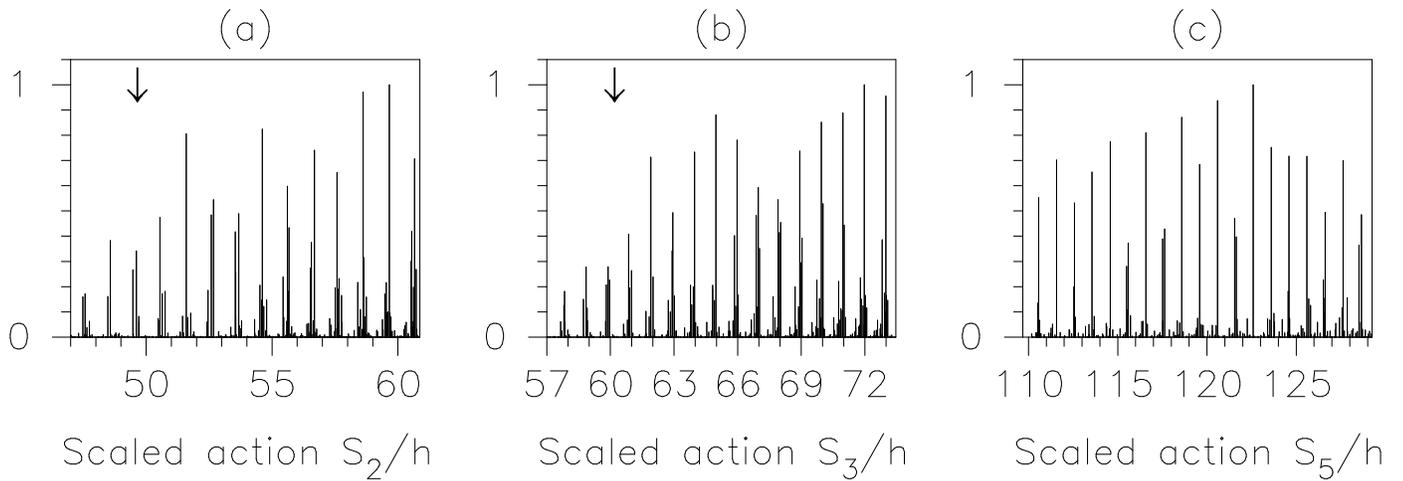}
\end{center}
\caption{
``Scar strength'' ( Husimi function at the location of the fixed point 
of the periodic orbit at the emitter barrier, calculated for 
the normal derivative of the wavefunction) vs. the 
scaled action $S(\varepsilon_n)/h$ of the corresponding orbit
($\varepsilon_n$ is the energy of the corresponding eigenstate)
for : (a) $(1,2)_2$ orbit, (b) $(1,3)^-$ orbit, (c) (1,5) orbit. 
The arrows indicate the values of $\beta$ for the tangent bifurcation, 
which give birth to the periodic orbits, the peaks of the scar strength 
below these values are due to the ``ghost effect'' 
\protect\cite{ref_ghost,ghost}. Scaled actions below the bifurcation points
were obtained by linear extrapolation of the 
(approximately linear) function $S(\varepsilon)/h$  
\label{overlap}
}
\end{figure}


\begin{references}
\bibitem{pitchfork} A subtle exception to this argument are orbits
which have only one collision with
the emitter, but generate two fixed points in the emitter SOS, corresponding
to the two different senses of traversing the orbit.  Such a $(1,2)$ orbit,
considered in ref. \cite{scars}, 
is created by a pitchfork bifurcation of the $(1,2)$ orbit considered here,
but it still has very small transverse velocity when it is near stability.
\end{references}
\end{document}